\documentclass[]{article}
\usepackage{graphicx}
\begin{document}
\title{Time-of-arrival probabilities and quantum measurements: II Application to tunneling times}
\author{Charis Anastopoulos\footnote{anastop@physics.upatras.gr}\\
{\small Department of Physics, University of Patras, 26500 Patras,
Greece}
\\
 and \\ Ntina Savvidou \footnote{ntina@imperial.ac.uk} \\
 {\small  Theoretical Physics Group, Imperial College, SW7 2BZ,
London, UK}} \maketitle

\begin{abstract}
We formulate quantum tunneling as a time-of-arrival problem: we
determine the detection probability for particles passing through a
barrier at a detector located a distance $L$ from the tunneling
region. For this purpose, we use a Positive-Operator-Valued-Measure
(POVM) for the time-of-arrival determined in \cite{AnSav06}. This
only depends on the initial state, the Hamiltonian and the location
of the detector. The POVM above provides a well-defined probability
density and an unambiguous interpretation of all quantities
involved. We demonstrate that for a  class of localized initial
states, the detection probability allows for an identification of
tunneling time with the classic phase time. We also establish limits
to the definability of tunneling time.

We then generalize these results to a sequential measurement set-up:
the phase space properties of the particles are determined by an
unsharp sampling before their attempt to cross the barrier. For such
measurements the tunneling time is defined as a genuine observable.
This allows us to  construct a probability distribution for its
values that is definable for all initial states and potentials. We
also identify a regime, in which these probabilities correspond to a
tunneling-time operator.

\end{abstract}

\renewcommand {\thesection}{\arabic{section}}
\renewcommand {\theequation}{\thesection. \arabic{equation}}
\let \ssection = \section \renewcommand{\section}{\setcounter{equation}{0} \ssection}

\section{Introduction}

This paper is a continuation of Ref (\cite{AnSav06}), in which  a
procedure was sketched for the construction of a
Positive-Operator-Valued Measure (POVM) for the time-of-arrival for
a particle described by a Hamiltonian $\hat{H}$. Here, we extend
this POVM to cover the case of particles tunneling through a
barrier. This procedure allows us to provide an unambiguous
determination for the tunneling-time as it can be measured in
time-of-arrival type of measurements.

Quantum tunneling refers to the escape of a particle from a region
 through a potential barrier, whose peak corresponds to an energy  higher
than that carried by the the particles. There are two important
questions (relevant to experiments) that can be asked in this
regard. The first is, how long does it take a particle to cross the
barrier (i.e. what is the tunneling time?). The second is, what is
the law that determines the rate of the particle's escape through
the barrier? In this paper, we develop a formalism that provides an
answer to these questions and we apply it to the first one. The
issue of the decay probability will be taken up in \cite{An07b}.

The issue of tunneling time has received substantial attention in
the literature, especially after the 1980's--see the reviews
\cite{reviews-tt, Sokol-book}. The reason is that there is an
abundance of candidates and a diversity of viewpoints with no clear
consensus. There are roughly three classes of approaches: (i) Wave
packet methods: one follows the particle's wave packet across the
barrier and determines the tunneling time through a "delay in
propagation" \cite{phasetime, wpm}, (ii) one defines suitable
variables for the particle's paths and one obtains a probability
distribution (or an average) for the transversal time corresponding
to each path. These paths can be constructed either through
path-integral methods \cite{Sokol-book, SokBa87, pathint}, through
Bohmian mechanics \cite{bohmian}, or through Wigner functions
\cite{wf}, and (iii) the use of an observable for time: this can
take the form of an additional variable playing the role of a clock
\cite{clock1, Larmor}, or of a formal time operator \cite{timeop}.
In general, these methods lead to inequivalent definitions and
values for the tunneling time.

\subsection{Our approach}

The basic feature of our approach to this problem is its operational
character. We identify the tunneling time by constructing
probabilities for the outcome of specific measurements. We assume
that the quantum system is prepared in an initial state $\psi(0)$,
which is localized in a region on one side of a potential barrier
that extends in a {\em microscopic} region. At the other side of the
barrier and a {\em macroscopic} distance $L$ away from
it\footnote{We explain in section 2.3 the sense in which we employ
the word "macroscopic".}, we place a particle detector, which
records the arrival of particles. Using an external clock to keep
track of the time $t$ for the recorder's clicks, we construct a
probability distribution $p(t)$ for the time of arrival. The fact
that the detector is a classical macroscopic object and that it lies
at a macroscopic distance away from the barrier allows one to state
(using classical language) that the detected particles must have
passed through the barrier (quantum effects like a particle crossing
the barrier and then backtracking are negligible). Hence, at the
observational level, the probability $p(t)$ contains all information
about the temporal behavior for the ensemble of particles: all
probabilistic quantities referring to tunneling can be reconstructed
from it.

 With the considerations above, both
problems of determining the tunneling time and the escape
probability as a function of time (see \cite{An07b}) are mapped to
the single problem of determining the time-of-arrival at the
detector's location for an ensemble of particles described by the
wave function $\psi_0$ at $t= 0$ and evolving under a Hamiltonian
with a potential term. To solve this problem, we elaborate on the
result of \cite{AnSav06}, namely the construction of a Positive
Operator Valued Measure (POVM) for the time-of-arrival for particles
for a generic Hamiltonian $\hat{H}$--see \cite{Davies} and
\cite{BLM96} for definition, properties and interpretation of POVMs.
 This POVM provides a unique
determination of the probability distribution $p(t)$ for the
time-of-arrival. It is important to emphasize that by construction
$p(t)$ is {\em linear} with respect to the initial density matrix,
positive-definite,  normalized (when the alternative of
non-detection is also taken into account) and a genuine density with
respect to time.

Since our results depend on the POVM for the time-of-arrival
constructed in \cite{AnSav06}, we review here the basic physical
considerations involved in its construction. The technical aspect,
namely the construction of this POVM for the problem at hand is
undertaken in Sec. 2.

The POVM of \cite{AnSav06} involves no structures other than the
ones of standard quantum mechanics: the Hamiltonian, the initial
state and the location of the recording device. It also involves a
smearing function with respect to time, but we employ it in the
regime in which the results are independent of such a choice. The
first step in the derivation arises from the remark that the notion
of arrival-time is well-defined when one considers {\em histories}
for a physical system (both in classical and in quantum
probability). We assume that the detector lies at $x = L$ and that
the initial state is localized in the region $I = \{x, x < L\}$.
Moreover, we assume a discretization $t_0, t_1, t_2, \ldots, t_n$ of
a time interval $[0, T]$. One asks at any instant $t_i$ of time,
whether the particle lies in region I or in region $II = \{x, x>
L\}$. The set of all possible successive alternatives forms a
Boolean algebra. The key point is that one can construct a
subalgebra of events labeled by the time of first crossing (together
with the event of no crossing), namely by the first instant $t_i$
that the particle is found in region $II$. This implies that
propositions about the time-of-arrival have a well-defined algebraic
structure, which is compatible with the Hilbert space description of
quantum mechanics.  The algebra of propositions for the
time-of-arrival is a special case of the so-called spacetime
coarse-grainings \cite{Har, scc} that have been studied within the
consistent histories approach to quantum mechanics \cite{Gri84,
Omn8894, GeHa9093, Har93a}.

The  construction above takes place at the discrete-time level. One
should then implement the continuum limit within the quantum
mechanical formalism. The problem is that there is no proper
continuous limit if one works at the level of probabilities (for the
same mathematical reason that leads to the quantum Zeno effect).
However, {\em there is} a proper continuum limit for this algebra if
one works at the level of amplitudes. More specifically, one can
implement the continuous limit at the level of the decoherence
functional, an object introduced in the consistent histories
approach.

The decoherence functional is a hermitian, bilinear functional on
the space of histories that contains all probability and phase
information for the histories of the system\footnote{Alternatively,
it can be viewed as a generating functional for all possible
temporal correlation functions of the system \cite{Ana01, Ana03}.}.
The restriction of the decoherence functional to the algebra of
propositions about the time-of-arrival effectively yields a
hermitian function $\rho(t, t')$ which is a density with respect to
both of its arguments.

The decoherence functional contains sufficient information for the
construction of POVMs for measurements that involve variables that
refer to more than one instant of time. This has been established
for sequential measurements \cite{Ana06} and for time-extended
measurements \cite{AnSav07}. In these cases one can compare the
results to ones obtained from single-time quantum mechanics, but for
the time-of-arrival, there is no analogous construction without the
use of histories. Nonetheless, the method provides a definition of
POVM for the time-of-arrival through a suitable smearing of the
diagonal elements of the decoherence functional. For a free
particle, this reproduces Kijowski 's POVM \cite{Kij74} in the
appropriate regime.

The important point in the procedure above is that the POVM of
\cite{AnSav06} is valid for a generic Hamiltonian. The time
parameter entering the POVM is the external Newtonian time and the
identification of the time-of-arrival is done through purely
kinematical arguments. Hence, this result can also be applied to the
specific Hamiltonian operators that are relevant to tunneling. This
is the content of Sec. 2.

Summarizing, there are three basic features in our approach: a) the
reformulation of tunneling as a time-of-arrival problem, b) the use
of POVMs for the determination of the probabilities for the
tunneling particles, and c) the basic ideas of the histories
approach that enable us to construct a suitable POVM.

\subsection{Relation to other approaches}

There are some common points and some points of divergence with
previous work on the tunneling time issue. Yamada has employed the
decoherence functional showing that different definitions of
tunneling time correspond to different definitions of the
alternatives for the `paths' considered in the definition
\cite{Yam04}. The construction of the decoherence functional is
different from ours in one respect: the (coarse-grained) histories
 we consider refer to the paths' first crossing of the
surface $x = L$, which lies a macroscopic distance away from the
barrier. In \cite{Yam04}, the histories refer to the crossing of
{\em the barrier} and the ambiguity in the definition of the
tunneling time reflects the inability to decide which of all
possible spacetime coarse-grainings provides the true measure of
tunneling time. This is due to the fact that quantum `mechanical'
paths may cross and then reenter the barrier region. In our case,
this is not an issue. The detector is far away from the barrier
region (at a macroscopic distance $L$) and the probability that a
particle crossing $x = L$ would ever backtrack to the barrier is
practically zero. Another difference is that Yamada argues within
the context of the decoherent histories programme that deals with
closed systems \cite{Yam99}. While we employ the methods and (many)
conceptual tools of consistent histories, our approach is strictly
operational within the Copenhagen interpretation. The probabilities
we construct refer to measurement outcomes in a statistical
ensemble. The decoherence functional is only used as a mathematical
object that allows us to construct a POVM and the particle crossing
of the surface $x = L$ is viewed as corresponding to an
(irreversible) act of measurement by a device located there.

The fact that the measurement of the particle takes place far away
from the barrier region suggests that our results should be
compatible with the asymptotic analysis of wave packets. Indeed, as
we shall see, our expression for tunneling time (whenever this can
be defined) corresponds to the classic Bohm-Wigner phase time
\cite{phasetime}. However, the methodology is different: we do not
identify time through the peak $x(t)$ of the wave-packet (or through
its center-of-mass), but the probability distribution for the
detection time is obtained from a POVM that is defined for all
possible initial states. Unlike time of detection, a sharp
definition for the tunneling time is only possible for initial
states characterized by a strong peak in their momentum
distribution. However, the generality of our construction allows us
to fully specify the limits in the definability of tunneling time.

From the technical point of view, our approach  has more in common
  with the second class of proposals we mentioned in the beginning:
  time being identified at the kinematical level from the properties
of `paths'. In particular, the formalism bears substantial
resemblance to the Feynman path integral derivation of tunneling
times of Sokolovski and Baskin \cite{SokBa87}. However, our boundary
conditions are different, and more importantly the probabilities we
obtain arise from  proper probability densities with respect to
time. While the time-averaged quantities in \cite{SokBa87} are
linear with respect to a restricted propagator, such propagators
appear in a quadratic form in our expression for the probability. It
was argued extensively in \cite{AnSav06} that this is necessary, in
order to obtain a genuine probability density in a way that respects
the convexity of the space of quantum states. The present
construction also shares these properties and this implies that the
issue of complex tunneling times does not arise.

\subsection{Our results }

In Sec.3, we apply the POVM we constructed in Sec. 2 to a simple
case of a particle in one-dimension. We consider  a potential
barrier $V(x)$, which takes non-zero values only in a bounded
(microscopic) region of width $d$ around $x = 0$. A wave packet
approaches the barrier from the negative real axis, while the
detection of the particle takes place at $x = L >> a$.

We assume that the initial wave -function is well localized in
position and in momentum (e.g. a coherent state). In addition, we
require that $\sigma/k_0 << 1$, where $k_0$ the mean momentum and
$\sigma$ the momentum spread of the initial state. We then find (for
a rather general regime for the values of the parameters
characterizing the system) that the probability distribution for the
time of arrival is sharply peaked around a value $t_m$.

From this probability distribution, we  identify the delay due to
the presence of the barrier as the difference between the time $t_m$
and the time it would take a classical particle of momentum $k_0$ to
travel from the center of the initial wavepacket to the location of
the detector. This `delay time' equals
\begin{eqnarray}
t_d = \frac{M}{k_0} \left(\frac{\partial T_k}{\partial k} \right)_{k
= k_0},
\end{eqnarray}

where $T_k$ is the transmission amplitude corresponding to the
potential $V(x)$. The delay time may be negative: the tunneling time
is obtained as the sum of $t_d$ with the time it would take the
particle to cross the barrier: it coincides with the classic phase
time.

Note that the delay and tunneling times defined this way are {\em
not}  observables of the system: they cannot be defined for a
generic initial state, but only for states well localized in
momentum and for values of the parameters that lead to a probability
distribution $p(t)$ characterized by a sharp peak. In this case, one
can use classical arguments for their definition. However, if either
the initial distribution has a substantial momentum spread, or if
$p(t)$ exhibits a more complex structure, there is no unambiguous
way to define tunneling time. While the value for the
time-of-arrival is a genuine observable (a random variable on the
sample space of the POVM), the tunneling time as we define it here
requires the knowledge of the corresponding time-of-arrival for a
free particle: and this cannot be defined, unless the initial value
of momentum is known\footnote{Strictly speaking, the above
definition of tunneling time involves counterfactual reasoning.
However, in the operational setting we consider here this is not a
problem, as long as we keep in mind that the tunneling time
(whenever it can be defined) is a `property' of the ensemble of
detected particles and not of any individual one.}.

Hence, in this approach the tunneling time is a {\em parameter} of
the detection probability. It can only be identified for specific
initial states, and not for any state, because its definition
involves a correspondence argument to classical physics.

In Sec. 5, we propose a generalization of the results above that
leads to a definition of the tunneling time as a genuine random
variable. The idea is to consider a sequential measurement set-up:
the phase space properties of the particle are determined through an
unsharp phase space sampling before this attempts to cross the
barrier, {\em and then} the time-of-arrival for the particles that
crossed the barrier is measured. The sample space corresponding to
such sequential measurements accommodates the definition of the
tunneling time as a genuine quantum observable and it allows us to
construct a marginal POVM that provides its probabilities for a
generic initial state. In a specific regime, this POVM becomes
independent of the details of the first measurement: as such it
defines an ideal probability distribution for the delay and the
tunneling times: this distribution suggests a definition for a
delay-time and for a tunneling-time operator.

\section{The general probability measure}
In this section, we review the construction of the POVM for the
time-of-arrival in \cite{AnSav06}, and we extend it for the case
relevant to tunneling.

\subsection{The histories formalism}

The POVM of \cite{AnSav06} is constructed using some notions of
quantum mechanical histories, as they appear in the consistent
histories approach to quantum theory of Griffiths \cite{Gri84},
Omn\'es \cite{Omn8894}, Gell-Mann and Hartle \cite{GeHa9093,
Har93a}. We should note however that these objects are used in the
present context differently, namely in an operational approach to
quantum theory--see \cite{Ana03, Ana06}

A   history  intuitively corresponds to a sequence of properties of
the physical system at successive instants of time. A discrete-time
history $\alpha$ is then represented by a string $\hat{P}_{t_1},
\hat{P}_{t_2}, \ldots \hat{P}_{t_n}$ of projectors, each labeled by
an instant  of time. From them, one can construct the class operator
\begin{equation}
\hat{C}_{\alpha} = \hat{U}^{\dagger}(t_1) \hat{P}_{t_1} \hat{U}(t_1)
\ldots \hat{U}^{\dagger} (t_n) \hat{P}_{t_n} \hat{U}(t_n)
\end{equation}
where $\hat{U}(s) = e^{-i\hat{H}s}$ is the time-evolution operator.
A candidate probability for the realisation of this history is
\begin{equation}
p(\alpha) = Tr \left( \hat{C}_{\alpha}^{\dagger}\hat{\rho}_0
\hat{C}_{\alpha} \right), \label{decfundef}
\end{equation}
 where $\hat{\rho}_0$ is the density matrix describing the system at time $t = 0 $.

However, the expression above does not define a probability measure
in the space of all histories, because the Kolmogorov additivity
condition cannot
 be satisfied: if $\alpha$ and $\beta$ are exclusive histories, and $\alpha \vee \beta$
denotes their conjunction as propositions, then it is not true that
\begin{equation}
p(\alpha \vee \beta ) = p(\alpha) + p(\beta) .
\end{equation}
 The
histories formulation of quantum mechanics does not, therefore,
enjoy the status of a genuine
 probability theory on the space of all histories.

However,  an
 additive probability measure {\it is} definable, when we restrict to
particular  sets of histories.
 These are called {\it consistent sets}. They are more conveniently
defined through the introduction of a new object: the decoherence
functional. This is a complex-valued function of a pair of histories
given by
\begin{equation}
d(\alpha, \beta) = Tr \left( \hat{C}_{\beta}^{\dagger} \hat{\rho}_0
\hat{C}_{\alpha} \right).
\end{equation}
A set of exclusive and exhaustive alternatives is called consistent,
if for all  pairs of different histories $\alpha$ and $\beta$, we
have $
 Re \; d(\alpha, \beta) = 0$.
In this case, one can use equation (\ref{decfundef}) to assign a
probability measure to this set.

\subsection{Time-of-arrival histories}

Using the histories formalism we construct a decoherence functional
for time-of-arrival histories with $N$ time steps $t_1, t_2, \ldots
t_N$(discrete-time). The reason for this construction is that the
decoherence functional has a good continuous time limit (unlike the
probabilities for histories).

We consider a particle in one dimension for concreteness, even
though the results obtained here only use abstract Hilbert space
operators and hold  more generally. We split the line into the
interval $(- \infty, L]$ and the interval $[L, \infty)$. Let
$\hat{P}_-$ and $\hat{P}_+$ be the corresponding projectors. Our aim
is to identify histories that correspond to the statement that the
particle crossed from the $-$ region to the $+$ region during a
particular time step.  If we assume that at $t = 0$ the particle
lies at the $-$ region then it is easy to verify that the history
\begin{eqnarray}
\alpha_m := (\hat{P}_-, t_1; \hat{P}_-, t_2; \ldots, \hat{P}_-, t_m;
\hat{P}_+, t_{m+1}; 1, t_{m+2}; \ldots 1, t_{N})
\end{eqnarray}

 corresponds to  the
proposition that the particle crossed $x=L$ for the first time
between the $m$-th and the $m+1$-th time step. The sequence
$\bar{\alpha} = (\hat{P}_-, t_1; \hat{P}_-, t_2; \ldots,  \hat{P}_-,
t_{m}; \ldots \hat{P}_-, t_{N})$ corresponds to the proposition that
the particle did not cross $x = L$ within the $n$- time steps.

The set of histories $\alpha_m$ together with $\bar{\alpha}$ is
exhaustive and exclusive (a sublattice of the lattice of history
propositions)--see also \cite{Har, Hal95}. The decoherence
functional is then defined on this set of histories: it is a
hermitian bilinear functional on a sample space consisting of the
points $(t_1, \ldots, t_n)$ together with the point $N$
corresponding to no crossing
\begin{eqnarray}
d(t_n, t_m) &=& d(\alpha_n, \alpha_m) \\
d(N, t_n ) &=& d(\bar{\alpha}, \alpha_n) \\
d(N,N) &=& d(\bar{\alpha}, \bar{\alpha}).
\end{eqnarray}

We next consider
  two discretisations $\{t_0 =0, t_1, t_2, \ldots t_N = T\}$ and $ \{t'_0
=0, t'_1, t'_2, \ldots t'_{N'} = T \}$ of the time interval $[0, T]$
with time-step $\delta t = T/N$, and $\delta t' = T/N'$. We
construct the decoherence functional $d([t, t+\delta t], [t',
t'+\delta t'])$,  where $ n = t N/T$ and $m = t' N'/T$. This reads
\begin{eqnarray}
d([t, t+\delta t], [t', t'+\delta t']) = Tr \left( \hat{\rho}_0
[e^{i \hat{H} \delta t'} \hat{P}_-]^n e^{i \hat{H}\delta t'}
\hat{P}_+ \right. \nonumber \\
\left. \times e^{i \hat{H}(t'-t)} \hat{P}_+ e^{-i \hat{H} \delta t}
[\hat{P}_- e^{-i \hat{H} \delta t}]^m \right).
\end{eqnarray}
We then take the limit $N, N' \rightarrow \infty$, while keeping $t$
and $t'$ fixed. Assuming that $\rho_0$ lies within the range of
$\hat{P}_-$, i.e. $\hat{P}_- \hat{\rho}_0 \hat{P}_- = \hat{\rho}_0$
we obtain
\begin{eqnarray}
d([t, t+\delta t], [t', t'+\delta t']) = \delta t \delta t' Tr
\left(   e^{i \hat{H}(t'-t)} \hat{P}_+ \hat{H} \hat{P}_- \hat{C}_{t}
\hat{\rho}_0 \hat{C}^{\dagger}_{t'} \hat{P}_- \hat{H} \hat{P}_+
\right),
\end{eqnarray}
where $\hat{C}_t = \lim_{n \rightarrow \infty} (\hat{P}_- e^{-i
\hat{H} t/n} \hat{P_-})^n$.  Writing
\begin{eqnarray}
\rho(t,t') = Tr \left(   e^{i \hat{H}(t'-t)} \hat{P}_+ \hat{H}
\hat{P}_- \hat{C}^{\dagger}_{t'} \hat{\rho}_0 \hat{C}_{t} \hat{P}_-
\hat{H} \hat{P}_+  \right) \label{densitydecf}
\end{eqnarray}
we see that the decoherence functional defines a complex-valued
density on $[0,T] \times [0,T]$. The additivity of the decoherence
functional (which reflects the additivity of quantum mechanical
amplitudes) allows us to obtain a continuum limit, something that
could not be done if we worked at the level of probabilities.

\subsection{The tunneling Hamiltonian}

For the simple case of a particle at a line with Hamiltonian
$\hat{H} = \frac{\hat{p}^2}{2 M} + V(\hat{x})$, where the potential
$V(x)$ is bounded from below, we employ a result in \cite{Har,
Facchi} that the restricted propagator $\hat{C}_t$ is obtained from
the Hamiltonian $\hat{H}$ by imposing Dirichlet boundary conditions
at $x = L$. If we also denote by $G_0(x,x'|t)$ the full propagator
in the position basis (corresponding to $e^{-i \hat{H}t}$), we
obtain
\begin{eqnarray}
\rho(t,t') &=& \frac{1}{4 M^2}\partial_x (\hat{C}_{t'} \psi_0)^*(L)
\partial_x(\hat{C}_t \psi_0)(L)
G_0(L,L|t-t') \label{basic}
\end{eqnarray}
where $\hat{\rho}_0 = | \psi_0 \rangle \langle \psi_0 |$, with
$\psi_0$ having support for $x < L$.

We now specialize to a case relevant for tunneling. We assume that
the potential is short-range: it is significantly different from
zero only in a region of width $d$ around $x = 0$. The distance $L$
is macroscopic, while $a$ is microscopic. This means that in the
neighborhood of  $x = L$ the propagator is effectively that of a
free particle. Hence, we can substitute $G_0(L,L|t'-t)$ in Eq.
(\ref{basic}) with the corresponding expression for the free
particle
\begin{eqnarray}
G(L,L|t) = \left( \frac{M}{2 \pi i t} \right)^{1/2}. \label{freep}
\end{eqnarray}

The considerations above also specify the range of values for $L$
that are relevant to our problem. The first condition on $L$ is that
the propagator may be substituted by that of the free particle, as
in Eq. (\ref{freep}). The second is that $L$ is sufficiently far
away from the tunneling region so that the probability amplitude of
a particle backtracking to the barrier region from $L$ is
practically zero. Physically one expects that this is the case for
all initial states $\psi_0$ for which the position spread $\Delta
q(t)$ remains at all times much smaller than $L$. Clearly, with the
considerations above it is not necessary that $L$ is a macroscopic
distance in the literal sense of the word: the requirement that $L$
be macroscopic is a sufficient but not a necessary condition.

We next consider the Hamiltonian $\hat{H}_D$ that is obtained from
the original Hamiltonian $\hat{H}$ by imposing Dirichlet boundary
conditions at $x=L$. We distinguish two cases: (i) if $x$ takes
value in the half-line, the spectrum of $\hat{H}_D$ is expected to
be discrete; (ii) if  $x$ takes values in the full real axis, at
least the positive energy spectrum will be continuous. (We restrict
to Hamiltonians having this property.) Either way, for $x
>> a$, $V(x) = 0$ and the solution of the Schr\"odinger equation
$\hat{H}_D \psi_E(x) = E \psi_E(x)$ with Dirichlet boundary
conditions is proportional to $\sin k(L -x)$, where $k =
(2ME)^{1/2}$. We choose to label the eigenstates of $\hat{H}_D$ by
$k$, namely we write $|k\rangle_D$ as a solution to the equation
\begin{eqnarray}
\hat{H} |k \rangle_D = \frac{k^2}{2M} | k \rangle_D,
\end{eqnarray}
with Dirichlet boundary conditions.

Normalizing $|k \rangle_D$ so that
\begin{eqnarray}
{}_D\langle k| k' \rangle_D = \delta (k, k'),
\end{eqnarray}
(and similarly in the discrete-spectrum case) we write
\begin{eqnarray}
 \langle x|k\rangle_D = D_k \sin k (L-x),
\end{eqnarray}
where the form of the normalization factor $D_k$ is specified the
Hamiltonian's (generalized) eigenstates.

For the study of tunneling, we assume that the initial state of the
system has support only in the positive energy spectrum of
$\hat{H}$. Hence,
\begin{eqnarray}
\langle x|\hat{C}_t |\psi_0 \rangle = \sum_k e^{-ik^2t/2M} D_k \sin
k(L-x) c_k,
\end{eqnarray}
where $c_k = {}_D\langle k|\psi_0 \rangle$ and $\sum_k$ denotes the
integration with respect to the spectral measure of $\hat{H}_D$.
Substituting into Eq. (\ref{basic}), we obtain
\begin{eqnarray}
\rho(t,t') = \frac{1}{4M \sqrt{2 \pi i M (t- t')}} \sum_{kk'} D_k
D_{k'}^* c_k c^*_{k'} kk' e^{-i\frac{k^2t-k'^2t'}{2M}}. \label{ro}
\end{eqnarray}

\subsection{Construction of the POVM}

The decoherence functional contains sufficient information for the
construction of POVMs for the probabilities of measurement outcomes
for magnitudes that have an explicit time-dependence. In particular,
the probabilities for the measurement outcomes for single-time,
sequential and extended-in-time measurements (obtained through the
standard formalism) can be identified with suitable diagonal
elements of the decoherence functional--see \cite{Ana06,AnSav07}. In
other words, one can define POVMs by suitable smearing of the
decoherence functional and in the cases above, these POVMs coincide
with ones obtained from the standard methods in quantum measurement
theory. In the case of the time-of-arrival there is no analogous
expression obtained from standard methods. However, the smeared form
of the decoherence functional still defines a POVM, and the main
assumption in this paper is that this POVM yields the correct
probabilities.

 With this assumption, we obtain the following probability density
for the time-interval $[0, T]$

\begin{eqnarray}
p^{\tau}(t) = \int_0^{T} ds \int_0^{T} ds' \sqrt{f^{\tau}(t,s)}
\sqrt{f^{\tau}(t,s')} \rho(s, s'), \label{ppp}
\end{eqnarray}

here $f_{\tau}(s, s')$ is a family of smeared delta functions
$f_{\tau}(s, s')$ characterized by the parameter $\tau$. The
functions $f_{\tau}$ satisfy the following property
\begin{eqnarray}
\int_0^T ds f^{\tau}(s, s') = \chi_{[0, T]}(s'),
\end{eqnarray}
where $\chi_{[0, T]}$ is the characteristic function of the interval
$[0, T]$: $\chi_{[0, T]}(s) = 1$ if $s \in [0, T]$, and $\chi_{[0,
T]} (s) = 0$ otherwise. The functions $f^{\tau}$  incorporate
specific features of the instrument that records particles crossing
the surface $x = L$.

Essentially, the key assumption in our approach (stated above) is
that the functions $f^{\tau}$ appearing in the definition of
(\ref{ppp}) are analogous to the smearing functions that appear in
the definition of POVMs for usual observables (i.e. ones other than
the time of arrival). In \cite{AnSav06}, we showed that this
assumption leads for the case of free particles to Kijowski's POVM
\cite{Kij74}.

The decoherence functional satisfies an hermiticity condition
$\rho(s, s') = \rho^*(s', s)$, which together with the positivity
condition for its diagonal elements
\begin{eqnarray}
\int_a^b ds \int_a^b \rho(s, s') \geq 0
\end{eqnarray}

guarantees that $p^{\tau}(t)$ is positive-definite for all values of
$t$.

The density (\ref{ppp}) is linear with respect to the initial
density matrix. Together with  the probability of no-detection
\begin{eqnarray}
p^{\tau}(N) = 1 - \int_0^T ds p^{\tau}(s)
\end{eqnarray}
they define a POVM $\hat{\Pi}$ on the space $[0, T] \cup \{N\}$.
This POVM describes the time of detection of a particle by an
instrument located at $x = L$.

In this paper, we will be interested in taking $T \rightarrow
\infty$, i.e. taking  $t \in [0, \infty)$. It is convenient to work
with Gaussian smearing functions
\begin{eqnarray}
f^{\tau}(s,s') = \frac{1}{\sqrt{2 \pi} \tau} e^{ - \frac{(s -
s')^2}{2 \tau^2}}. \label{Gauss}
\end{eqnarray}
However, these Gaussians are smeared delta-functions with respect to
the whole real axis and {\em not} with respect  to the time-interval
$[0, \infty)$. To remedy this problem, we note that by Eq.
(\ref{ro}), $\rho(-s, -s') = \rho^*(s,s') = \rho(s', s)$ and that
the probability (\ref{ppp}) is symmetric to an exchange of $s$ and
$s'$. We also note that the contribution of terms that mix positive
and negative $s$  are significant only at times $|t|$ of order
$\tau$. Hence, the probability (\ref{ppp}) with the Gaussian
(\ref{Gauss}) is substituted in place of $f^{\tau}$ and the
integration limits taken from $- \infty$ to $\infty$,  is twice the
probability density that is defined with an integral over the
positive half-axis. Hence, the use of Gaussian smearing functions
only involves dividing $p^{\tau}(t)$ in (\ref{ppp}) by a factor of 2
(for times $t>> \tau$).

Inserting (\ref{Gauss} into (\ref{ppp}), we change variables to $u =
\frac{1}{2} (s+s')$ and $v = s-s'$ noting that
\begin{eqnarray}
\sqrt{f^{\tau}(t,s)} \sqrt{f^{\tau}(t,s')} = f_{\tau} (u - t) e^{-
\frac{v^2}{8 \tau^2}}.
\end{eqnarray}
We substitute in the integration $f_{\tau}(u-t)$ with a delta
function $\delta(u-t)$. We then obtain
\begin{eqnarray}
p^{\tau}(t) = \frac{1}{8M \sqrt{2 \pi  M }} \sum_{kk'} D_k D_{k'}^*
c_k c^*_{k'} kk' e^{-i\frac{k^2- k'^2}{2M}t} \; R\left(\frac{k^2 +
k'^2}{4M}\right), \label{fullppp}
\end{eqnarray}
where
\begin{eqnarray}
R(\epsilon) = \int_{-\infty}^{\infty} dv \frac{e^{-v^2/8\tau^2 - i
\epsilon v}}{\sqrt{iv}} = 2 \sqrt{\tau} \int_0^{\infty} dy
\frac{e^{-y^2/2} [\cos (2 \epsilon \tau y) + \sin (2 \epsilon \tau
y)]}{\sqrt{y}}.
\end{eqnarray}
At the limit of $\epsilon \tau >> 1$, i.e. if the detection time is
much larger than $\epsilon^{-1}$ \footnote{This condition is valid
if the mean energy of the initial state is much larger than the
energy uncertainty, and it is accurate for all times $t >> \tau$.}
\begin{eqnarray}
\int_0^{\infty}dy \;  \frac{ e^{ - \frac{y^2}{2}} [ \cos (2 \epsilon
\tau y) + \sin(2 \epsilon \tau y)]}{\sqrt{y}} \simeq
\sqrt{\frac{\pi}{\epsilon \tau}}.
\end{eqnarray}
Hence, $R(\epsilon) = 2 \sqrt{\pi/\epsilon}$. It follows that
\begin{eqnarray}
p(t) = \frac{1}{ 2\sqrt{2 }M} \sum_{kk'} D_k D_{k}^* c_k c^*_{k'}
\frac{kk'}{\sqrt{k^2 +k'^2}} e^{-i \frac{k^2 - k'^2}{2M} t}.
\label{probab}
\end{eqnarray}

The probability for the time-of-arrival then becomes independent of
the parameter $\tau$, and it is expressed solely in terms of the
system's Hamiltonian, the initial state and the value of $L$.

 Eq. (\ref{probab})  is
simplified if the spread $\Delta k$ of the initial state
$|\psi_0\rangle$ ( $\hat{k} = \sqrt{2M\hat{H}_D}$) is much smaller
than the corresponding mean value $\bar{k}$: in this case, $k^2 +
k'^2 \simeq 2 kk'$, hence
\begin{eqnarray}
p(t) =  \left| \sum_k D_k c_k \sqrt{\frac{k}{4M}}
e^{-ik^2t/2M}\right|^2.
\end{eqnarray}

It was shown in \cite{AnSav06} that for the test case of a free
particle (, in which $D_k = (2 \pi)^{-1/2}$)
 the probability distribution above reproduces the one of
Kijowski \cite{Kij74}.

\section{The detection probability}

In this section, we use the probability density (\ref{probab}) in a
specific context that allows us to determine a magnitude that
corresponds to the time the particle spends in the forbidden region.
In effect, we identify  tunneling-time  by the delay caused by the
presence of the barrier to the particles' time-of-arrival (see Sec.
4). This turns out to be the same definition as the one employed in
the methods involving the wave packet analysis. However, we do not
identify any specific features of the wave-packet (these objects
have no natural probabilistic or operational interpretation in
quantum mechanics), but we work directly at the level of measurement
outcomes, namely the probability distribution for the
time-of-arrival.

We  consider the simplest possible example of a particle tunneling
through a potential barrier. We assume that the potential $V(x) \geq
0 $ takes non-zero values in a region of width $d$ around $x=0$. Let
$V_0$ be the maximum value of this potential. In classical mechanics
no particle with energy $E < V_0$ can cross the barrier, hence the
probability of detection at $x = L$ is zero at all times. We next
consider this problem in quantum theory.

Eq. (\ref{probab}) involves the eigenstates of the Hamiltonian with
Dirichlet boundary conditions at $x = L$. Since $x \in (-\infty,
\infty)$, the spectrum of the Dirichlet Hamiltonian is continuous.
The summation over $k$ is then substituted by an integral
$\int_0^{\infty} dk$.

The first step is to construct the generalized eigenstates of the
Hamiltonian with Dirichlet boundary conditions. To do so, we first
study the solutions to the Schr\"odinger equation
\begin{eqnarray}
-\frac{1}{2M} \partial^2_x u(x) + V(x) u(x) = \frac{k^2}{2M} u(x).
\end{eqnarray}
There are two linearly independent solutions for each value of $k$.
It will necessary to construct an orthonormal basis of generalized
eigenstates from these solutions. We pick one class of solutions
$u_k^+(x)$ that correspond to a particle propagating from $-\infty$
and scattering on the potential
\begin{eqnarray}
u_k^+(x) = \left\{ \begin{array}{c} A^+_k \left( e^{ikx} + R^+_k
e^{-ikx} \right) \; \; x < -d/2 \\ A_k^+ T^+_k e^{ikx}\; \; \; \; \;
\; \; \; \; \; \; \; \; \; \; x
>d/2
\end{array} \right.,
\end{eqnarray}
where $R_k^+$ and $T_k^+$  is the reflection and transmission
coefficient respectively, while $A_k^+$ is a normalization factor so
that $\int dx \bar{u}_k^+(x) u_k^+(x) = \delta(k-k')$. Let $u^-_k$
be a normalized  linearly independent solution that satisfies $\int
dx \bar{u}_k^+(x) u^-_k(x) = 0$. Its form will be the following
\begin{eqnarray}
u_k^-(x) = \left\{ \begin{array}{c} A^-_k
\left( T^-_k e^{-ikx} + S_k e^{ikx} \right) \; \; x < -d/2 \\
A_k^- \left( e^{-ikx} + R^-_k e^{ikx} \right) \; \; x
>d/2
\end{array} \right.
\end{eqnarray}
Note that there is no reason for $u^-_k$ to have a physical
interpretation in terms of left-moving particles, and the labels
$T_k^-, R_k^-$ are chosen for convenience: they do not correspond to
a transmission and reflection coefficient of any short. We also note
that the coefficients in $u^+_k, u^-_k$ are not independent. For any
two solutions $\psi, \phi$ to the Schr\"odinger equation with the
same energy, the Wronskian $\psi' \phi - \phi' \psi$ must be
$x$-independent. This yields the following conditions
\begin{eqnarray}
T^+_k = T^-_k - S_k \bar{R}^+_k \\ \label{cond1}
S_k = \bar{T}^+_k R^-_k + T^-_k \bar{R}^+_k \\
|T^+_k|^2 + |R^+_k|^2 = 1 \\
|T^-_k|^2 + |R^-_k|^2 = 1 + |S_k|^2. \label{cond4}
\end{eqnarray}

To impose the Dirichlet boundary conditions on these solutions, we
take a linear combination $v_k(x)$ of $u_k^+(x)$ and $u_k^-(x)$ and
require that $v_k(L) = 0 $. This yields
\begin{eqnarray}
v_k(x) = C_k \left[ A_k^- (1 + R_k^- e^{2ikL}) u_k^+(x) - A_k^+
T_k^+ e^{2ikL} u_k^-(x) \right],
\end{eqnarray}
where
\begin{eqnarray}
C_k = \frac{1}{\sqrt{|A_k^-|^2  |1 + R_k^- e^{2ikL}|^2 + |A_k^+|^2
|T_k^+|^2}}
\end{eqnarray}
is a normalization constant chosen so that $\int dx \bar{v}_k(x)
v_{k'}(x) dx = \delta(k-k')$.

For $x > d/2$, we obtain
\begin{eqnarray}
v_k(x) = - 2i  C_k A_k^- A_k^+ T^+_k e^{ikL} \sin k(L-x).
\end{eqnarray}
Hence,
\begin{eqnarray}
D_k = - 2i  C_k A_k^- A_k^+ T^+_k e^{ikL}
\end{eqnarray}

We now consider a Gaussian initial state $\psi_0$ centered around
$x_0 < -d/2$ and having mean momentum $k_0 > 0$
\begin{eqnarray}
\psi_0(x) = \frac{1}{(2 \pi \delta^2)^{1/4}} e^{- \frac{(x-x_0)^2}{4
\delta^2} + i k_0 x}, \label{initial}
\end{eqnarray}
where $\delta$ is the spread in position and we assume that $ \delta
<< |x_0 + d/2|$ so that the initial state does not overlap with the
region where the potential is non-zero. In this region,
\begin{eqnarray}
v_k(x) = C_k A_k^- A_k^+  \left[  (1 + R_k^- e^{2ikL} - T_k^+ S_k
e^{2ikL}) e^{ikx} \right. \nonumber \\
\left.  + (R_k^+ R_k^- e^{2ikL} - T_k^+ T_k^- e^{2ikL}) e^{-ikx}
\right].
\end{eqnarray}

The coefficients $c_k = {}_D\langle k| \psi_0 \rangle $ are then
given by
\begin{eqnarray}
c_k =  \bar{C}_k \bar{A}_k^- \bar{A}_k^+ \left[ 1 + (\bar{R}_k^- -
\bar{T}_k^+ \bar{S}_k) e^{-2ikL} \right] \frac{1}{(2 \pi
\sigma^2)^{1/4}} e^{ - \frac{(k- k_0)^2}{4 \sigma^2} - ix_0 (k +
x_0)},
\end{eqnarray}
where we set $\sigma = (2 \delta)^{-1}$ the momentum spread. The
assumption that $\sigma/k_0 << 1$  allowed us to drop a term of
order $e^{-k_0^2/4\sigma^2}$.

The probability for the time-of-arrival at $x = L$ is then given by
$p(t) = |z(t)|^2$, where
\begin{eqnarray}
z(t) = \int_0^{\infty} dk \; B_k \; e^{ikL} \; \frac{1}{(2 \pi
\sigma^2)^{1/4}} e^{ - \frac{(k- k_0)^2}{4 \sigma^2} - ix_0 (k -
k_0)} \sqrt{\frac{k}{4M}} e^{-ik^2t/2M}. \label{z}
\end{eqnarray}
In (\ref{z}) we defined
\begin{eqnarray}
B_k = -2i \sqrt{2 \pi}|C_k|^2 |A_k^-|^2 |A_k^+|^2 \left[ 1 +
(\bar{R}_k^- - \bar{T}_k^+ \bar{S}_k) e^{-2ikL} \right] T_k^+.
\label{beta}
\end{eqnarray}
Since $\sigma/k_0 << 1$, we can expand $B_k$ around its value at $k
= k_0$. As a term $\sqrt{k}$ also appears in the integral outside
the exponential, we expand together
\begin{eqnarray}
\sqrt{k} B_k \simeq \sqrt{k_0} B_{k_0} e^{(\xi_{k_0} + i
\lambda_{k_0})(k-k_0)}, \label{expansion0}
\end{eqnarray}
where
\begin{eqnarray}
\xi_{k_0}  = \frac{1}{2k_0} + \left(\frac{\partial \log
|B_k|}{\partial k}\right)_{k = k_0}\\
\lambda_{k_0} = \left(\frac{\partial \arg[B_k]}{\partial
k}\right)_{k = k_0}.
\end{eqnarray}
Within the same approximation, we  take the limits of integration in
Eq. (\ref{z}) from $-\infty$ to $\infty$. We then obtain
\begin{eqnarray}
z(t) = B_{k_0} e^{ik_0L} \sqrt{\frac{k_0}{4M}} \frac{1}{(2 \pi
\sigma^2)^{1/4}} \nonumber \\
\times \int_{-\infty}^{\infty} dk e^{ - \frac{(k- k_0)^2}{4
\sigma^2} + i(|x_0| + L+ \lambda_{k_0} - i \xi_{k_0})  (k - k_0)}
 e^{-ik^2t/2M}.
\end{eqnarray}
The expression above involves a standard Gaussian integral. Its
evaluation gives
\begin{eqnarray}
z(t) = B_{k_0} e^{-i k_0^2t/2M+ ik_0L} \sqrt{\frac{k_0}{4M}} \frac{
(8 \pi \sigma^2)^{1/4}}{\sqrt{1 + 2 i t \sigma^2/M}} \nonumber \\
\times \exp \left[ - \sigma^2 \frac{ (|x_0| + L + \lambda_{k_0} -
\frac{k_0t}{M} - i \xi_{k_0})^2}{1 + 2it \sigma^2/M} \right].
\end{eqnarray}
Hence,
\begin{eqnarray}
p(t) = |z(t)|^2 = |B_{k_0}|^2 e^{2 \sigma^2 \xi_{k_0}^2}
\frac{k_0}{4M} \sqrt{\frac{8 \pi \sigma^2}{1 + 4 t^2 \sigma^4/M^2}}
\nonumber \\
\times \exp \left\{ - \frac{2 k_0^2 \sigma^2/M^2}{1 + 4 t^2
\sigma^4/M^2} \left[(1 + 2 \xi_{k_0} \sigma^2/k_0) t - \frac{M(|x_0|
+ L + \lambda_{k_0})}{k_0}\right]^2 \right\} \label{p1}
\end{eqnarray}

This expression is the probability distribution for the
time-of-arrival, as it would be measured by a device located at
distance $L$ from the barrier. In the following section, we analyze
its properties: in particular, we identify the delay caused by the
presence of the barrier.

\section{Delay-time and tunneling time}

\subsection{The identification of delay time}

For a sufficiently monochromatic wave packet ($\sigma/k_0
\rightarrow 0$), we assume that $\xi_{k_0} \sigma^2/k_0 <<1$, hence
Eq. (\ref{p1}) yields

\begin{eqnarray}
p(t) = |z(t)|^2 = |B_{k_0}|^2 e^{2 \sigma^2 \xi_{k_0}^2}
\frac{k_0}{4M} \sqrt{\frac{8 \pi \sigma^2}{1 + 4 t^2 \sigma^4/M^2}}
\nonumber \\
\times \exp \left\{ - \frac{2 k_0^2 \sigma^2/M^2}{1 + 4 t^2
\sigma^4/M^2} \left[ t - \frac{M(|x_0| + L +
\lambda_{k_0})}{k_0}\right]^2 \right\} \label{p2}
\end{eqnarray}

The term $1 +4 t^2 \sigma^4/M^2$ corresponds to the spread in the
particle's wave function due to time evolution. Since we want a
configuration in which the determination of time is as sharp as
possible, we assume that the value of $\sigma$ is so small that this
spread is negligible at the time $t_m = \frac{M(|x_0| + L +
\lambda_{k_0})}{k_0}$, namely that $t_m^2 \sigma^2/M << 1$. Then we
obtain

\begin{eqnarray}
p(t) = |B_{k_0}|^2 e^{2 \sigma^2 \xi_{k_0}^2} \frac{k_0}{4M} \sqrt{8
\pi \sigma^2} \exp \left\{- \frac{2 k_0^2 \sigma^2}{M^2} \left[ t -
\frac{M(|x_0| + L + \lambda_{k_0})}{k_0}\right]^2 \right\}.
\label{p3}
\end{eqnarray}

Then $t = t_m$ is a sharp peak for the mean value of the
time-of-detection. A classical particle (or in quantum theory a
narrow wavepacket) that starts from $x_0$  with momentum $k_0$ {\em
in absence of the potential barrier} will arrive at $x = L$ at
(average) time $t_0 = M \frac{|x_0| + L}{k_0}$. Hence, the barrier
causes a `delay' $t_d = t_m - t_0$ to the time-of-arrival (of the
particles that are not reflected)

\begin{eqnarray}
t_d = M \lambda_{k_0}/k_0.
\end{eqnarray}
The presence of the barrier has increased the effective length that
has to be traversed by the particle by a factor of $\lambda_{k_0}$.
In fact, $\lambda_{k_0}$ may be negative: the time it takes the
particle to cross the forbidden region of the barrier is $ t_{tun} =
M (\lambda_{k_0} + d_{k_0})/k_0$, where $d_{k_0} = x_2(k_0) -
x_1(k_0) \geq 0$, where $x_{1,2}(k_0)$  are the points that
determine the forbidden region: they are respectively the lowest-
and highest-valued solutions of the equation $\frac{k_0^2}{2M} =
V(x)$. The total tunneling time has to be positive, but it is not
necessary that it is larger than the time $Md_{k_0}/k_0$ that the
forbidden region is traversed  by a classical free particle.

We next calculate $\lambda_{k_0}$ in terms of the absorbtion and
reflection coefficients corresponding to the potential $V(x)$. From
Eq. (\ref{beta}) we see that the only term contributing to a phase
in $B_k$ is the product $\left[ 1 + f_k e^{-2ikL} \right] T_k^+ $,
where $f_k = (\bar{R}_k^- - \bar{T}_k^+ \bar{S}_k)$. We then obtain
\begin{eqnarray}
\lambda_{k_0} = Im \left( \frac{\partial \log T_k^+}{\partial k}
\right)_{k = k_0} + Im \frac{f'_{k_0} - 2iL f_{k_0}}{1 + f_k e^{-2i
k_0 L}} e^{-2i k_0L}. \label{lll}
\end{eqnarray}

The second term in the right-hand-side of (\ref{lll}) oscillates
very fast with $L$, because $L$ is much larger than the de-Broglie
wavelength $2 \pi/k_0$ of the particle. These oscillations are an
artifact of our modeling the detection process by a crossing of the
sharply defined surface $x = L$. In a realistic detection scheme the
particle detection  cannot take place with an accuracy grater than
their de Broglie wavelength. For this reason, we can formally
average $L$ within a region of size $l <<L$.
 Indeed, using a Gaussian smearing
function $\rho(L) = (\pi l^2)^{-1/2} e^{ - (L - L_0)^2/l^2}$ we
obtain a suppression factor of order $e^{ - k_0^2 l^2} << 1$ for the
oscillating terms.

Hence,  the effective tunneling time is
\begin{eqnarray}
t_{tun} = \frac{M d_{k_0}}{k_0} + \frac{M}{k_0} Im \left(
\frac{\partial \log T_k^+}{\partial k} \right)_{k = k_0},
\end{eqnarray}
i.e. we recover the  expression for the Bohm-Wigner phase time
\cite{phasetime}. It is important to emphasize that this derivation
did not employ any characteristics of the wave-packets (e.g. the
trajectory followed by their peak, or their `center-of-mass'). It is
a natural {\em operational} definition at the level of the
probability density that corresponds to the measurement outcomes.

Note that a precise treatment involves smearing the probability
function $p(t)$ of (\ref{p1}). The only $L$-dependent objects that
appear in this equation are the term $B_{k_0}$ and the Gaussian
exponential. If $\frac{1}{\sigma} >> l$, the effect of smearing is
to substitute $L$ by the mean value $L_0$: the expression is not
affected.  The effect of smearing on $B_{k_0}$ is to suppress the
oscillations; it leads to an effective expression $\tilde{B}_{k_0}$
\begin{eqnarray}
\tilde{B}_{k_0} = -2i \sqrt{2 \pi}  \frac{|A_{k_0}^-|^2
|A_{k_0}^+|^2}{|A_{k_0}^-|^2  (1 + |R_{k_0}^-|^2) + |A_{k_0}^+|^2
|T_{k_0}^+|^2} T_{k_0}^+. \label{btil}
\end{eqnarray}

Note that to a first (very rough) approximation, $|A_{k_0}^{\pm}|$
can be taken equal to $(2 \pi)^{-1/2}$, i.e. the value taken if the
contribution of the region with no zero potential is considered to
be negligible. Then
\begin{eqnarray}
\tilde{B}_{k_0} \simeq - \frac{i}{\sqrt{2 \pi}} T_{k_0}^+
\label{btil2}
\end{eqnarray}

Before continuing, we summarize the approximations involved in the
results we obtained in this section. Eq. (\ref{z}) only involves the
assumption that $\sigma/k_0 <<1$. Eq. (\ref{p1}) involves the
additional assumption that the function $\log B_k$ varies slowly
around $k = k_0$ so that it is sufficient to keep the first order in
its Taylor expansion. This approximation amounts to the condition
$|\frac{B_{k_0}''}{B_{k_0}'} - \frac{B_{k_0}'}{B_{k_0}}| \sigma
<<1$. Eq. (\ref{p2}) involves the additional assumption that
$\xi_{k_0} \sigma^2/k_0 << 1$. Finally, Eq. (\ref{p3}) involves the
assumption that $t_m^2 \sigma^2/M << 1$. This implies that $L$
cannot be too large, because the spread of the wave function due to
the free propagation will induce a large uncertainty in the
determination of tunneling time.

\subsection{Special cases}

\paragraph{Parity-invariant potentials.} The expression for the mode
functions and for $B_k$ simplifies greatly if the potential is
invariant under parity, namely if $V(x) = V(-x)$. This implies that
the eigenstate $u_k^-(x)$ can be identified with the parity
transform of $u_k^+(x)$. Hence $S_k = 0$, $T_k^+ = T_k^- := T_k$,
$R^+_k = R^-_k := R_k$ and $A_k^+ = A_k^- := A_k$. We then obtain,
\begin{eqnarray}
B_k = - 2i \sqrt{2 \pi}\frac{|A_k|^2}{|1 +R_k e^{2ikL}|^2 + |T_k|^2}
[1 + \bar{R}_k e^{-2ikL}] T_k
\end{eqnarray}

\paragraph{The square potential barrier}  We apply our results to the simplest example of a square
potential barrier: $V(x) = V_0$ for $x \in [-d/2, d/2]$. Defining
$\gamma_k = \sqrt{2MV_0 - k^2}$, we obtain the following values for
the coefficients $T_k, R_k$
\begin{eqnarray}
T_k = \frac{2 k}{\gamma_k} e^{-ikd} \frac{2k \gamma_k [2 k \gamma_k
\cosh \gamma_k d - i  (\gamma^2_k - k^2) \sinh \gamma_k d]}
{4 k^2 \gamma_k^2 + (\gamma_k^2 + k^2) \sinh^2 \gamma_k d} \\
R_k = -i e^{-ikd}  \frac{(\gamma^2 + k^2) [2 k \gamma \cosh \gamma_k
d - i  (\gamma_k^2 - k^2) \sinh \gamma_k d]}{4 k^2 \gamma_k^2 +
(\gamma_k^2 + k^2) \sinh^2 \gamma_k d} .
\end{eqnarray}

There are two limits, in which the results are particularly simple.
The limit of a long barrier $\gamma_k d >> 1$, for which

\begin{eqnarray}
T_k &\simeq& e^{-ikd} e^{-\gamma_k d} \frac{4 k
\gamma_k}{(\gamma_k^2 + k^2)^2} [2k \gamma_k - i (\gamma_k^2 - k^2)]
\label{tklong}
\\
R_k &\simeq& e^{-ikd} \frac{-(\gamma_k^2 - k^2) + ik\gamma_k}{4
\gamma_k^2 } \label{rklong}
\end{eqnarray}
In this limit, the parameter $\lambda_{k_0}$ is
\begin{eqnarray}
\lambda_{k_0} = -d + \frac{2}{\gamma_{k_0}} ,
\end{eqnarray}
i.e. it takes negative values (since $\gamma_{k_0} d >> 1$). The
tunneling time is therefore $t_{tun} = \frac{2M}{ \gamma_{k_0}
k_0}$.

The other limit is that of the delta function (very short) barrier.
It is obtained by letting $V_0 \rightarrow \infty$ and $d
\rightarrow 0$ such that $V_0 d$ is a constant (we denote this
constant as $\kappa/M$). At this limit, $\gamma_k d \simeq
\sqrt{\kappa d}$ and
\begin{eqnarray}
T_k = \frac{1}{1 + i \kappa/k} \label{tkdelta} \\
R_k = \frac{1}{1 + i k/\kappa}. \label{rkdelta}
\end{eqnarray}
Hence,
\begin{eqnarray}
\lambda_{k_0} = \frac{\kappa}{k_0^2 + \kappa^2}.
\end{eqnarray}
Since $d = 0$ the tunneling time is $t_{tun} = M
\frac{\kappa/k_0}{k_0^2 + \kappa^2}$.

\subsection{Comments}

\subsubsection{Domain of validity}

 It is important to emphasize that our identification of a
tunneling time $t_{tun}$ relies on the fact that the probability of
detection has a unique sharp maximum at a specific moment of time.
This is only possible for specific initial states. For example, it
is easy to demonstrate that a superposition of Gaussians centered at
different values of momentum will lead to a probability distribution
with an oscillating behavior. While there is still a mean detection
time, we cannot read from it a time delay for the particle, because
the momentum uncertainty does not allow one to specify uniquely a
corresponding time for free particle evolution. Hence, the tunneling
time is not a proper observable (i.e. a random variable on the
sample space upon which the POVM is defined) in our description: it
is only a parameter that appears in the detection probability for a
class of initial states, which has an intuitive interpretation in
terms of classical concepts.

The fact that the concept of tunneling time has a restricted domain
of validity is highlighted by another point. We saw that for a long
square potential  the tunneling time equals $t_{tun} = \frac{2M}{
\gamma_{k_0} k_0}$. If $d$ is very large, the  condition
$\gamma_{k_0} d
>> 1$ can be satisfied even if $\gamma_{k_0}$ takes very small
values, i.e. if the particle's mean energy $\frac{k_0^2}{2M}$ is
very close to $V_0$. Hence, it is in principle possible to construct
configurations, in which $t_{tun}$ is arbitrarily small: the
effective `velocity' $d/t_{tun}$ in the crossing of the barrier is
then super-luminal. This is a well known effect in tunneling (the
Hartman effect \cite{Hartmann}). A full treatment in the present
context involves the consideration of relativistic systems--this we
will undertake in future work. Here, we only note that the regime of
very large values for $d /t_{tun}$, (very small values for
$\gamma_{k_0}$) is one for which the approximation involved in Eq.
(\ref{expansion0}) fails. The tunneling probability increases
rapidly in this regime and one would have to include further terms
in the expansion of $\log B_k$, which would lead to a substantially
deformed probability distribution $p(t)$ with no clear peak. The
definition of $t_{tun}$ would then be highly problematic, and so
would be the notion of a mean velocity in the tunneling region.

\subsubsection{Uncertainty in the specification of tunneling time}

The uncertainty in the determination of the peak in the probability
distribution (\ref{p3}) is $\frac{M}{k_0 \sigma}$. In order for the
delay time $\frac{M \lambda_{k_0}}{k_0}$ to be distinguishable (if
we ignore all other sources of uncertainty) it is necessary that
$\sigma |\lambda_{k_0}|
>> 1$. In order for the  tunneling time to be
distinguishable, it is also necessary to take into account the
uncertainty in the quantity $\frac{M d_{k_0}}{k_0}$. To leading
order in $\sigma$ this equals $a_{k_0} \sigma$, where
\begin{eqnarray}
a_{k_0} = \frac{k_0}{M} \left(\frac{1}{V'[x_2(k_0)]} -
\frac{1}{V'[x_1(k_0)]} \right) - \frac{M d_{k_0}}{k_0^2}.
\end{eqnarray}

The overall uncertainty in the determination of the tunneling time
$t_{tun} = M(\lambda_{k_0} + d_{k_0})/k_0$ is of the order
\begin{eqnarray}
\frac{M }{k_0 \sigma} + |a_{k_0}| \sigma.
\end{eqnarray}
This expression is bounded from below by $2 \sqrt{M k_0 a_{k_0}}$.
Hence, a necessary condition for  tunneling time $t_{tun}$ to be
distinguishable is
\begin{eqnarray}
t_{tun} >> \sqrt{M  |a_{k_0}|/k_0}.
\end{eqnarray}
We note that for a parity symmetric potential $a_{k_0} = - M
d_{k_0}/k_0^2$, hence the condition becomes $t_{tun} >> M \sqrt{
d_{k_0}/k_0}$. For the long square barrier, this implies that
\begin{eqnarray}
\frac{ \gamma_{k_0}^2 d }{k_0} << 1.
\end{eqnarray}
This condition can only be satisfied if $\gamma_{k_0}/k_0 << 1$.
This is inadmissible, because the expansion (\ref{expansion0}) is
not adequate in this regime. Hence, for the long square barrier the
operational definition of the tunneling time is not meaningful. On
the other hand, there is no problem in the short barrier  limit ($d
\rightarrow 0$).

\subsubsection{The dependence on $L$ }

Finally, we comment on the assumption that $L >>d$.  The
consideration of a detector at a macroscopic distance away from the
barrier region greatly simplifies our results: it leads to an
expression for the tunneling time, which essentially coincides with
the results of the asymptotic analysis of the wave packets. This
assumption enters at two steps. First, in the construction of the
POVM, we assume that $L$ is sufficiently removed from the barrier
region, so that the value for the particle's propagator at $x = L$
can be substituted by the corresponding value for the free particle.
This condition is satisfied exactly if the corresponding Hamiltonian
has no (generalized) eigenstates with an asymptotic behavior that
does not correspond to that of a free particle (e.g. negative energy
states). This is the case we considered in this section. Hence, the
only place where the assumption of large $L$ enters in a non-trivial
way in the construction, is when we smear the probability
distribution in order to remove the contribution of the terms
oscillating as $e^{ik_0L}$. This implies that (at least formally),
the expression (\ref{lll}) for the parameter $\lambda_{k_0}$ is
valid for all values of $L$ such that the first condition stated
above holds. We therefore obtain an expression for the tunneling
time, even if the detector is located near the tunneling region.
Clearly, this will have a very sensitive dependence on $L$, because
the presence of the detector close to the barrier affects the
configuration of the system. Note however that this result is rather
formal, since it involves the idealization of the detection process
by the crossing of the sharply defined surface $x = L$. In a
realistic treatment the detailed physics of the detector are
expected to influence the tunnelling time.

For example, for a parity symmetric potential ($S_{k} = 0 $), we
obtain the following expression for the parameter $\lambda_{k_0}$,
\begin{eqnarray}
\lambda_{k_0} = \theta'_{k_0}  \frac{r_{k_0}(2L + \theta'_{k_0}) [1
+ \cos (2 k_0 L +\theta_{k_0})] + r'_{k_0}  \sin (2k_0 L +
\theta_{k_0})}{1 + r_{k_0}^2 + 2 r_{k_0} \cos (2 k_0 L +
\theta_{k_0})}, \; \; \; \; \;
\end{eqnarray}
where we wrote $R_{k} = r_k e^{i \theta_k}$ and the prime denotes
differentiation with respect to $k$.

\section{A POVM for the tunneling time through sequential
measurements}

We saw in the last section that the determination of tunneling time
through the time-of-arrival probability is only meaningful for a
specific class of initial states, because the delay time is not a
proper random variable on the sample space of the POVM. It depends
on the particle's initial momentum (and position) and as such it
cannot be inferred unless both the initial state and the detection
probability have very sharp maxima.

However, this problem can be alleviated if we make a change in the
experimental set-up, namely if we consider that a measurement of
momentum takes place before any recording of the time-of-arrival. In
effect, if one considers sequential measurements, it is possible to
construct a POVM for which the tunneling time is a genuine random
variable and no mixed classical-quantum arguments are needed for its
identification.

The procedure is the following. Let $\hat{Q}(x,k)$ be a POVM for
unsharp phase space measurements. Let us also assume that the
corresponding device is placed at the left-hand-side of the barrier;
we perform an unsharp phase space measurement to  any particle that
moves towards the barrier that allows us to determine unsharp values
for its position $q$ and momentum  $p$. The measurement is assumed
to be non-destructive, hence the particles continue their motion,
some of them cross the barrier and they are detected at distance $L$
away. In other words, we have a sequential measurement: first an
unsharp phase space measurement and then a time-of-arrival
measurement. For each particle, the outcomes of this sequential
measurement is encoded in the three numbers $(x, k, t)$ that span a
sample space $\Omega$.

The key point is that from the knowledge of $\hat{Q}$ and
$\hat{\Pi}$ (the time-of-arrival POVM), it is possible to construct
a POVM $\hat{E}$ on $\Omega$. The procedure is standard, see
\cite{Ana06} for a detailed analysis. The POVM $\hat{E}$ consists of
the positive operators
\begin{eqnarray}
\hat{E}(t, x, k) = \sqrt{\hat{Q}}(x, k) \hat{\Pi}(t)
\sqrt{\hat{Q}}(x, k),
\end{eqnarray}
and of the positive operator
\begin{eqnarray}
\hat{E}(N, x, k) = \sqrt{\hat{Q}}(x, k) \hat{\Pi}(N)
\sqrt{\hat{Q}}(x, k),
\end{eqnarray}
that corresponds to a phase space measurement and then no detection.
By construction it satisfies
\begin{eqnarray}
\int{dx dk}{2 \pi} \left(\int_0^{\infty} dt \hat{E}(t, x, k) +
\hat{E}(N, x, k) \right) = 1.
\end{eqnarray}

For an initial state $\hat{\rho}_0$, the joint probability density
on the sample space $\Omega$ is given by
\begin{eqnarray}
P(t, x, k) = Tr \left( \hat{\rho}_0 \hat{E}(t, x, k) \right).
\end{eqnarray}

The key benefit in the consideration of such a POVM is that the
delay-time
\begin{eqnarray}
t_d = t - \frac{M(L - x)}{k},
\end{eqnarray}
and the tunneling time
\begin{eqnarray}
t_{tun} = t_d + \frac{M d_k}{k},
\end{eqnarray}
are both random variables on the sample space $\Omega$. Hence, it is
possible to define a POVM on the space in which they take values. We
will do so after we construct explicitly the POVM $\hat{E}$.

We consider POVMs for the unsharp phase-space measurements of the
form
\begin{eqnarray}
\hat{Q}(x, k) = \int \frac{dk_0 dx_0}{2 \pi}  f(x - x_0, k - k_0)
|x_0, k_0 \rangle \langle x_0, k_0|,
\end{eqnarray}
where $|x_0, k_0 \rangle$ is the coherent state (\ref{initial}), and
$f$ is a positive-valued function that determines the phase space
resolution of the apparatus.  Since $\int \frac{dx dk}{2 \pi}
\hat{\Pi}(x, k) = 1$, it is necessary that the function $f$
satisfies
\begin{eqnarray}
\int \frac{dx dk}{2 \pi} f( x, k) = 1.
\end{eqnarray}

The minimum resolution measurements correspond to $f(x,k) = 2 \pi
\delta(x) \delta (k)$,  in which case $\hat{Q}(x, k) = |x k \rangle
\langle x k|$. For simplicity, we will consider minimum resolution
measurements in what follows.

We obtain the following probability density on $\Omega$
\begin{eqnarray}
P(t, x, k) = \langle x k|\hat{\rho}|x k \rangle \langle x
k|\hat{\Pi}(t) | x k \rangle.
\end{eqnarray}
We note that $\langle x k|\hat{\Pi}(t) | x k \rangle$ equals the
probability density $p(t)$ of Eq. (\ref{p1}). We write this as
$p_{x, k}(t)$, in order to express its dependence on the initial
state [$k = k_0$ and $x  = x_0$ in Eq. (\ref{p1})]. We then obtain
\begin{eqnarray}
 P(t, x, k) = \langle x k|\hat{\rho}|x k \rangle  p_{x,k}(t). \label{povmseq}
\end{eqnarray}

We next change variables in (\ref{povmseq}) from $t$ to the delay
time $t_d$. We note that on the full sample space, the relation
between $t_d$ and $t$ is not one-to-one. First, the random variable
$t_d$ takes values in the whole real axis, while $t$ only on the
positive real axis. It is therefore convenient to define the
probability $P(t, x, k)$ for $t$ running to all reals. This involves
defining $p_{x,k}(t)$ for all $t \in {\bf R}$; we saw in Sec. 2 that
this is obtained by doubling the values of $p_{x,k}(t)$ for $t \in
[0, \infty)$. With $t$ defined over all reals, we note that for each
value of $t$, one obtains the same value for $t_d$ {\em twice},
since $t_d$ is the same at points $(t, x, p)$ and $(t, 2L-x, -p)$.
We perform the change of variables taking the facts above into
account, and then we integrate over
 $x$ and $k$, in order to obtain a
marginal probability distribution over $t_d$
\begin{eqnarray}
P_d(t_d) = 4 \int \frac{dx  dk}{2 \pi}  \langle x k|\hat{\rho}|x k
\rangle  p_{x,k}(t_d + \frac{L - x}{k}).
\end{eqnarray}
The same procedure leads to a marginal probability distribution for
the tunneling time
\begin{eqnarray}
P_{tun}(t_{tun}) = 4 \int \frac{dx  dk}{2 \pi}  \langle x
k|\hat{\rho}|x k \rangle p_{x,k}(t_{tun} + \frac{L - x + d_k}{k}).
\end{eqnarray}

The two equations above are completely general, and they hold
without any approximations. They simplify significantly if we assume
that for all values of $k$ in the support of the initial state, the
following two conditions hold: (i) $p_{x, k}(t)$ is appreciably
different from zero only for times $t$ such that
 $t^2
\sigma^2/M << 1$, and (ii) $\sigma \xi_k << 1$. The dependence on
$x$ of $p_{x,k}$ is then absorbed in the definition of the variable
$t_d$, and we obtain
\begin{eqnarray}
P_d(t_d) = \sqrt{8 \pi \sigma^2} \int dk \langle k|\hat{\rho_0} |k
\rangle |\tilde{B}_{k}|^2 \frac{|k|}{M}  \exp \left\{- \frac{2 k^2
\sigma^2}{M^2} \left[ t_d - \frac{M \lambda_k}{k}\right]^2 \right\}.
\label{pd}
\end{eqnarray}
Similarly,
\begin{eqnarray}
P_{tun}(t_{tun} ) = \sqrt{8 \pi \sigma^2} \int dk \langle
k|\hat{\rho_0} |k \rangle |\tilde{B}_{k}|^2  \frac{|k|}{M} \exp
\left\{- \frac{2 k^2 \sigma^2}{M^2} \left[ t_{tun} - \frac{M
(\lambda_k + d_k)}{k}\right]^2 \right\}. \label{pt}
\end{eqnarray}
Note that neither $P_d$ nor $P_{tun}$ are normalized to unity. The
delay and tunneling times are only defined for the fraction of the
ensemble that corresponds to particles that have crossed the
barrier. To normalize it, we have to divide by the probability
corresponding to the detected particles $1 - Tr \left( \hat{\rho}_0
\hat{E}(N) \right)$.

Hence, we have constructed a positive definite probability density
for the delay and the tunneling times, which is valid for an
arbitrary initial state (with the restriction that its position
support lies on the left side of the barrier). This probability is
definable in the context of a sequential measurement: there is no
other way to define these times as quantum observables otherwise:
the definition in Sec.4 involved a mixture of quantum mechanics and
classical argumentation and was only meaningful for a specific class
of initial states.  We have to keep in mind though that the
experimental set-up for which these probabilities are valid involves
keeping track of the phase space properties of {\em individual
particles} and then comparing them with the registered arrival time.
It requires relatively precise measurements at a microscopic scale,
and it cannot be implemented when working with particle beams.

We should also note that both probabilities $P_d$ and $P_{tun}$ are
contextual, i.e. they depend strongly on specific features of the
apparatus that performs the phase space sampling. They both have a
strong dependence on the parameter $\sigma$, which defines the
family of coherent states: in the present context $\sigma$ is the
inherent uncertainty in the specification of momentum\footnote{For
the contextuality of sequential measurements, see the extended
discussion in \cite{Ana06}}. At the limit $\sigma \rightarrow 0$,
both (\ref{pd}) and (\ref{pt}) vanish. There is, however, a limit in
which the results become $\sigma$-independent. If the initial state
has support on values of $k$, such that the mean of the Gaussian in
either probability density is much larger than its spread, then we
can approximate it by a delta function. This condition implies
\begin{eqnarray}
\sigma |\lambda_k| >> 1, \label{conditiond}
\end{eqnarray}
for (\ref{pd}) and
\begin{eqnarray}
\sigma (\lambda_k + d_k) >> 1 \label{conditiont}
\end{eqnarray}
for (\ref{pt}).

At these regimes, we obtain
\begin{eqnarray}
P_d(t_d) = 2\pi \int dk  \langle k|\hat{\rho_0} |k \rangle
|\tilde{B}_{k}|^2 \delta (t_d - \frac{M \lambda_k }{k}), \label{pdid} \\
P_{tun}(t_{tun}) = 2 \pi \int dk  \langle k|\hat{\rho_0} |k \rangle
|\tilde{B}_{k}|^2 \delta (t_{tun} - \frac{M (\lambda_k + d_k) }{k}).
\label{ptid}
\end{eqnarray}
In other words, the values of $P_d(t_d)$ and of $P_{tun}(t_{tun})$
are determined by the value of the probability distribution of the
initial's state momentum at values of $k$ that are solutions of the
algebraic equations $t_d = \frac{M \lambda_k }{k}$ and $t_{tun} =
\frac{M (\lambda_k + d_k) }{k}$ respectively. These expressions for
the probability distribution are independent of the detailed
characteristics of the phase space POVM: they only depend on the
initial state and on the characteristics of the
potential\footnote{Recall that by virtue of smearing the value of
$L$, there is no $L$-dependence in $\tilde{B_k}$; hence the marginal
probability distributions are also $L$-independent.}. They can
therefore be considered as {\em ideal} distributions of delay and
tunneling times respectively that exhibit little sensitivity to the
measurement scheme employed for their determination.

We can further simplify the expressions for $P_d$ and $P_{tun}$
using the  estimation (\ref{btil2}) for $\tilde{B}_{k_0}$:
\begin{eqnarray}
P_d(t_d) =  \int dk  \langle k|\hat{\rho}_0 |k \rangle
|T_{k}|^2 \delta (t_d - \frac{M \lambda_k }{k}), \label{pdid2} \\
P_{tun}(t_{tun}) =  \int  dk  \langle k|\hat{\rho}_0 |k \rangle
|T_{k}|^2 \delta (t_{tun} - \frac{M (\lambda_k + d_k) }{k}).
\label{ptid2}
\end{eqnarray}

In effect, the probability for $t_d$ and $t_{tun}$ are defined from
the corresponding values of the momentum distribution weighted by
the transmission probability. Defining the functions $F_d(k) :=
\frac{M \lambda_k }{k}$ and $F_{tun}(k) := \frac{M (\lambda_k + d_k)
}{k}$, we see that the probabilities (\ref{pdid2}-\ref{ptid2}) are
obtainable from the operators $\hat{T}_{d} = F_{d}(\hat{p})$ and
$\hat{T}_{tun} = F_{tun}(\hat{p})$ ($\hat{p}$ is the momentum
operator) when these act on the state
\begin{eqnarray}
\hat{\rho}_{cross} = \int dk |T_k|^2 \hat{P}_k \hat{\rho}_0
\hat{P}_k, \hspace{2cm} \hat{P}_k = |k \rangle \langle k|
\end{eqnarray}
that describes the sub-ensemble of particles that have crossed the
barrier. One could therefore call $\hat{T}_d$ and $\hat{T}_{tun}$
time-delay and tunneling-time operators respectively\footnote{There
is an ambiguity in their definition at $k = 0$. However, this does
not affect the probabilities (\ref{pdid2}) and (\ref{ptid2}),
because $|T_{k = 0}| = 0$.}.

We end this section, by examining the domain of validity of
conditions (\ref{conditiont}) and (\ref{conditiond}) for the square
potential barrier. At the large barrier limit, they read
\begin{eqnarray}
\sigma |-d + \frac{2}{\gamma_k}| >> 1 \\
\sigma/\gamma_{k} >> 1.
\end{eqnarray}
They are satisfied if the position $\sigma^{-1}$ spread of the
coherent states is much smaller than the effective lengths
corresponding to delay and tunneling time respectively. For the
delta function barrier, these conditions imply
\begin{eqnarray}
\sigma \frac{\kappa}{k^2 + \kappa^2} >> 1,
\end{eqnarray}
which is only possible if $\kappa$ is extremely small (a rather
unphysical case).

We see therefore that the ideal probability distributions
(\ref{pdid}) and (\ref{ptid}) can only be obtained if the initial
phase space measurement has a resolution for position substantially
smaller than the dimensions of the barrier. This is a type of
measurement that is not explicitly forbidden by quantum mechanics,
but clearly it would be extremely difficult to achieve in practice.

\section{Conclusions}

We reformulated tunneling as a problem in the determination of
probability for the time-of-arrival. This allowed us to identify the
classic Bohm-Wigner time as the most suitable measure for the
tunneling time. However, this identification only holds for a
specific class of initial states and potentials; in other regimes,
there is no operational definition of the concept. There is one way
to go around this problem by considering a sequential measurement
set-up: we first measure the phase space properties of the particles
(before they attempt to cross the barrier) {\em and then} we
determine their times-of-arrival.  In this context, it is possible
to construct a probability measure for the tunneling time that is
valid for all initial states.

The key feature of our construction is that there is neither
interpretational nor probabilistic ambiguity. The probabilities we
derive are obtained through a POVM, hence they are always positive
and they respect the convexity of the space of quantum states. The
interpretation of these objects is concretely operational, in the
sense that it is tied to the statistics for the measurement of
particles' arrival times. Tunneling time is solely defined in terms
of the statistics of measurement outcomes.

In another paper \cite{An07b}, the POVM we constructed here will be
employed for the study of the decay probability of unstable states
through tunneling.

\section*{Acknowledgements}

N.S. acknowledges support from the EP/C517687 EPSRC grant.

\end{document}